\documentstyle[prc,aps,epsfig]{revtex}
\newcommand{\diff}{{\rm\,d}}                    
\newcommand{\ove}{\overline}                    
\def\fps@figure{htb}
\def\fps@table{htb}

\def\s{\mbox{\boldmath $s$}}

\def\N{\mbox{\boldmath $N$}}
\def\L{\mbox{\boldmath $L$}}                                  
\def\r{\mbox{\boldmath $r$}}

\def\p{\mbox{\boldmath $p$}}
\def\P{\mbox{\boldmath $P$}}
\def\q{\mbox{\boldmath $q$}}
\def\v{\mbox{\boldmath $v$}}
\def\k{\mbox{\boldmath $k$}}
\def\T{\mbox{\boldmath $T$}}
\def\e{\mbox{\boldmath $e$}}
\def\N{\mbox{\boldmath $N$}}
\def\J{\mbox{\boldmath $J$}}

\def\ss{\mbox{\boldmath $\sigma$}}

\def\dd{\mbox{\boldmath $\nabla$}}

\def\r{\mbox{\boldmath $r$}}
\def\pf{\mbox{\boldmath $p^{ \prime}$}}
\def\j{\mbox{$\jmath$}}
\mathchardef\varepsilon="010F
\mathchardef\epsilon="0122
\mathchardef\theta="0123
\mathchardef\vartheta="0112
\begin{document}
\draft
\title{Relativistic corrections in  ($\e,\e'\p$) knockout reactions}
\author{A.~Meucci, C.~Giusti, and F.~D.~Pacati}
\address{Dipartimento di Fisica Nucleare e Teorica dell'Universit\`a, 
Pavia\\
and Istituto Nazionale di Fisica Nucleare, Sezione di Pavia, Italy}
\date{\today}
\maketitle

\begin{abstract}

A consistent comparison between nonrelativistic and relativistic descriptions 
of the $(e,e'p)$ reaction is presented. 
We use the nonrelativistic {\tt DWEEPY} code and 
develop a fully relativistic model starting from the
effective Pauli reduction for the scattering state and the relativistic mean
field theory for the bound state. Results for the $^{16}$O$(e,e'p)$
differential cross section and structure functions are compared in various
kinematical conditions. A limit in energy of the validity of the nonrelativistic
approach is established.
The effects of spinor distortion and of the effective momentum approximation 
for the scattering state are discussed. A satisfactory agreement with data of 
differential cross sections, structure functions, and polarization observables 
is achieved.
\end{abstract}

\section{Introduction}

Exclusive ($e,e'p$) knockout reactions represent a very clean tool to explore
the single-particle (s.p.) aspects of nuclear structure revealing the properties 
of proton-hole states contained in the hole spectral 
function~\cite{FM,PR,Oxford,Kelly1}. 

Several high-resolution experiments were carried out at Saclay~\cite{FM,Mougey} 
and NIKHEF~\cite{NIKHEF}. The analysis of the missing energy and momentum 
dependence of the experimental cross sections allowed to assign specific 
quantum numbers and spectroscopic factors to the peaks in the energy spectrum. 
The calculations for this analysis were carried out with the program 
{\tt DWEEPY}~\cite{DWEEPY}, within the theoretical framework of a 
nonrelativistic distorted wave impulse approximation (DWIA), where final-state 
interactions (FSI) and Coulomb distortion of the electron wave functions are 
taken into account. Phenomenological ingredients were used to compute bound 
and scattering states. The outgoing nucleon scattering wave functions are 
eigenfunctions of an optical potential determined through a fit to elastic
nucleon-nucleus scattering data including cross sections and polarizations. 
Bound-state wave functions were calculated with a Woods-Saxon well, where the 
radius was determined to fit the experimental momentum distributions and the 
depth was adjusted to give the  experimentally observed separation energy of 
the bound final state. This theoretical approach was able to describe, with a 
high degree of accuracy, in a wide range of nuclei and in different kinematics, 
the shape of the experimental momentum distributions at missing-energy values 
corresponding to specific peaks in the energy spectrum. In order to reproduce 
the size of the experimental cross sections, the normalization of the 
bound-state wave function was fitted to the data and identified with the 
spectroscopic factor. These values, however, are smaller than those predicted 
by many-body theories. 

Similar models based on a fully relativistic DWIA (RDWIA) framework were 
developed in more recent years~\cite{Pick,Jin,Udias,Ud1,HP,Joha,Joha1}. In these
approaches the bound nucleons are described by s.p. Dirac wave functions in the
presence of scalar and vector potentials fitted to the ground-state properties
of the nucleus, and the scattering wave function is solution of the Dirac 
equation with relativistic optical potentials obtained by fitting elastic 
proton-nucleus scattering data. Some of these approaches include also an exact 
treatment of the Coulomb distortion of the electron waves~\cite{Jin,Udias}. Also 
RDWIA  analyses were able to give a good description of the experimental 
momentum distributions. Thus, the shape of the distributions in DWIA and RDWIA 
calculations are  similar, while the spectroscopic factors, obtained by scaling 
the calculated cross sections to the data, are in RDWIA about $10-20$\% larger 
than in DWIA analyses, and thus closer to theoretical predictions. The 
difference was attributed to the relativistic optical potential, which leads to 
stronger absorption. The nucleon-nucleus interaction exhibits characteristic 
non-localities which arise quite naturally in the Dirac approach and whose 
effect was not included in standard nonrelativistic DWIA calculations, but  
can be accounted for in the nonrelativistic treatment by a renormalization of 
the scattering wave function. The so-called Darwin normalization 
factor~\cite{Cannata}, that essentially changes the Schr\"odinger wave function 
into the upper component of the Dirac wave function, increases the absorption 
due to FSI and produces a quenching of the calculated cross section by about 
15\%, with a corresponding enhancement of the spectroscopic factor in agreement 
with the results obtained in RDWIA.  

New data have recently become available from TJNAF. The cross section for 
quasielastic $1p$-shell proton knockout has been measured and the response 
functions and the asymmetry have been extracted in the $^{16}$O($e,e'p$) 
reaction at four-momentum transfer squared $Q^2 = 0.8$ (GeV/$c$)$^2$ and energy 
transfer $\omega \sim  439$ MeV~\cite{Gao}. In the same kinematics also first 
polarization transfer measurements have been carried out for the 
$^{16}$O(${\vec e},e'{\vec p}$) reaction~\cite{Malov}. The polarization of the 
ejected proton in the $^{12}$C($e,e'{\vec p}$) reaction has been measured at 
Bates with $Q^2 = 0.5$ (GeV/$c$)$^2$ and outgoing-proton energy 
$T_{\mathrm p} = 274$ MeV~\cite{Woo}. 

The analysis of these new data in kinematic conditions unaccessible in previous
experiments, where $Q^2$ was less than 0.4 (GeV/$c$)$^2$ and $T_{\mathrm p}$ 
generally around 100 MeV, requires a theoretical treatment where all 
relativistic effects are carefully and consistently included. Indeed these 
recent data are well described by RDWIA calculations\cite{Joha1,Gao,Malov,Woo}.

Fully relativistic models based on the RDWIA have been developed by different
groups~\cite{Pick,Jin,Udias,Ud1,HP,Joha,Joha1}. It was shown in Ref.~\cite{HP} 
that the hadronic part of the relativistic transition amplitude can be written 
in terms of Schr\"odinger-like wave functions for bound and scattering states 
and of an effective nuclear current operator which contains the Dirac 
potentials. In this way, complications due to the solution of the Dirac equation 
can be avoided and spinor distortion can be accounted for solving a relativized 
Schr\"odinger equation, of the same type as that solved in ordinary 
nonrelativistic DWIA calculations with nonrelativistic or relativistic 
equivalent potentials, but Dirac scalar and vector potentials appear in the 
nuclear current operator. This so-called effective Pauli reduction does not 
represent an approximation. It is in principle exact and can be considered an 
alternative fully relativistic approach. 

This approach was adopted in Refs.~\cite{Kelly2,Kelly3} with an 
effective momentum approximation (EMA) to incorporate spinor distortion into the
effective current operator. In this approximation the momentum operators in the 
term $\ss \cdot \p$, appearing in the lower components of the Dirac spinor and 
in the effective current operator, are replaced by their asymptotic values. 
This model contains approximations, but accounts for all the main relativistic 
effects and is able to give a good description of the most recent experimental 
results. The spectroscopic factor extracted in comparison with $^{16}$O($e,e'p$)
data for $1p$-shell proton knockout is the same as in other RDWIA analyses, 
i.e. 0.7~\cite{Gao}. 

Both nonrelativistic DWIA and RDWIA models are able to describe, with a good 
degree of accuracy, the ($e,e'p$) data at low energies, but the nonrelativistic 
DWIA approach is more flexible. Some relativistic corrections can be included in 
a nonrelativistic treatment, but a fully relativistic model is needed for the 
analysis of the data at higher energies that are now becoming available. It is 
thus important to establish a clear relationship between the nonrelativistic 
DWIA treatment that was extensively used in the analysis of low-energy data and 
RDWIA treatments, in order to understand the relevance of genuine relativistic 
effects and the limit of validity of the nonrelativstic model. 

Relativistic effects as well as the differences between relativistic and
norelativistic calculations have been widely and carefully investigated in
different papers where RDWIA treatments have been developed. The differences,
however, were usually investigated starting from the basis of a relativistic 
model where terms corresponding to relativistic effects are cancelled, such as, 
for instance, the lower components in the Dirac spinor and the Darwin factor, or
where nonrelativistic approximations are included. Although very interesting, 
these investigations do not correspond to the result of a direct and consistent 
comparison between RDWIA and the DWIA calculations carried out with the program 
{\tt DWEEPY}, that was used in the analysis of low-energy data. In fact, 
{\tt DWEEPY} is based on a nonrelativistic treatment where some relativistic 
corrections are introduced: a relativistic kinematics is adopted and 
relativistic corrections at the lowest order in the inverse nucleon mass are 
included in the nuclear current operator~\cite{MVVH,GP}, which is derived from 
the Foldy-Wouthuysen transformation~\cite{FW}. The Darwin factor can be simply 
included in a nonrelativistic treatment~\cite{Cannata}, but it was not 
explicitly considered in the data analyses carried out with {\tt DWEEPY}.

Only indirect comparison between relativistic and norelativistic calculations 
can be obtained from the available data analyses carried out with {\tt DWEEPY} 
and in RDWIA. In fact, the two types of calculations make generally use of 
different optical potentials and bound state wave functions, and the difference 
due to the different theoretical ingredients cannot be attributed to relativity. 
This problem was already clear in previous investigations of relativistic 
effects. Various attempts were made to reproduce the conditions of a 
nonrelativistic calculation from a relativistic approach, but a consistent and 
direct comparison has not yet been achieved. 

The main aim of this paper is to make clear the relationship between the DWIA
approach in {\tt DWEEPY} and RDWIA treatments, and to investigate the relevance 
of genuine relativistic effects through a direct comparison between the results 
of the two calculations. A fully relativistic RDWIA treatment of the ($e,e'p$) 
knockout reaction has thus been developed.  The effective Pauli 
reduction~\cite{HP} has been adopted for the scattering state. For the bound 
state the numerical solution of the Dirac equation has been used, as in this
case it does not represent a too difficult problem. Various computer programs 
are in fact available, able to explain the global and s.p. properties of a 
nucleus within the framework of a relativistic mean-field theory. For the 
scattering state, where a partial-wave expansion is performed, the effective
Pauli reduction appears simpler and more flexible, and it is equivalent to the
solution of the Dirac equation. From this point of view, our relativistic
approach does not contain approximations, since the Schr\"odinger-like  equation 
is solved for each partial wave starting from a relativistic optical potential 
and without assuming the EMA prescription in the effective current operator. 

The numerical results of this RDWIA model have been compared with the results 
of {\tt DWEEPY}. A consistent comparison requires the same kinematical conditions and
the use of consistent theoretical ingredients in the two calculations. Thus, in 
{\tt DWEEPY} we have adopted for the bound state the normalized upper component 
of the Dirac spinor and for the scattering state the solution of the same 
Schr\"odinger-equivalent optical potential of the relativistic calculation. 

Different kinematics have been considered for the comparison, with the aim to
investigate the relevance of relativistic effects not included in {\tt DWEEPY} 
in different situations, at low energy, in the kinematical region where 
{\tt DWEEPY} was extensively and successfully applied, and at higher energy, 
where relativistic effects are more evident, in order to establish the limit 
of validity of the nonrelativistic approach. 

The relevant formalism is outlined in Sec.~II. Relativistic and 
nonrelativistic calculations of the cross section and response functions for 
the $^{16}$O($e,e'p$) reaction are compared in Sec.~III, where various 
relativistic effects are investigated. Even though the comparison with 
experimental data is not the main aim of this work, in Sec.~IV we check 
the reliability of our approach in comparison with data. Some conclusions are 
drawn in Sec.~V.

\section{Formalism}

\subsection{Relativistic current}

In the one-photon exchange approximation, where a photon is exchanged between 
the incident electron and the target nucleus, the coincidence cross section of
the ($e,e'p$) reaction is given by the contraction between the lepton tensor, 
dependent only on the electron variables and completely determined by QED, and 
the hadron tensor, whose components are given by bilinear products of the 
transition matrix elements of the nuclear current operator. According to the 
impulse approximation, in which only one nucleon of the target is involved in 
the reaction, the nuclear current is assumed to be a one-body operator.
 
In RDWIA the matrix elements of the nuclear current operator, i.e.
\begin{eqnarray}
J^{\mu } = \int \diff \r \ \ove \Psi _f(\r )\ \hat {\j }^{\mu }\ 
 \exp \{{\mathrm i}\,\q \cdot \r \} \ \Psi _i(\r ),
\label{eq.relcur}
\end{eqnarray} 
are calculated with relativistic wave functions for initial bound and final
scattering states. 

We choose the electromagnetic current operator corresponding to the $cc2$
definition of Ref.~\cite{deF}, i.e. 
\begin{eqnarray}
\hat {\j }^{\mu } = F_1(Q^2)\gamma ^{\mu }\ +\ {\mathrm i}\,
\frac {\kappa }{2M}F_2(Q^2)\ 
 \sigma ^{\mu \nu } q_{\nu }, 
\label{eq.cc2}
 \end{eqnarray}
where $q^{\nu}=(\omega,\q)$ is the four-momentum transfer, $Q^2=q^2-\omega^2$, 
$F_1$ and $F_2$ are Dirac and Pauli nucleon form factors, $\kappa$ is the 
anomalous part of the magnetic moment, and 
$\sigma ^{\mu \nu } = {\mathrm i}/2\ \big [\gamma ^{\mu },\gamma ^{\nu }\big ]$.
Current conservation is restored by replacing the longitudinal current by
\begin{equation}
J^{\mathrm L} = J^z = \frac {\omega }{q} J^0 .
\end{equation}  
In our reference frame the $z$-axis is along $\q$ and the $y$-axis parallel to
$\q \times \p'$.

The bound state wave function
\begin{eqnarray} 
\Psi _i = \pmatrix {u_i \cr v_i} \label{eq.bouwf}
\end{eqnarray} 
is obtained as the Dirac-Hartree solution from a relativistic Lagrangian with 
scalar and vector potentials. Several computer codes calculating the ground and 
excited state properties of nuclei are easily available in
literature~\cite{TIM,pvr}.

The ejectile wave function is obtained following the direct Pauli reduction 
method. It is well known that a Dirac spinor
\begin{eqnarray}
\Psi = \pmatrix { \Psi _+ \cr \Psi_-}
\end{eqnarray}
can be written in terms of its positive energy component $\Psi _+$ as
\begin{eqnarray}
\Psi = \pmatrix { \Psi _+ \cr \frac {\ss \cdot \p}{E+M+S-V}\ \Psi _+},
\label{eq.poscom} 
\end{eqnarray}
where $S=S(r)$ and $V=V(r)$ are the scalar and vector potentials for the
 nucleon with energy $E$. The upper component $\Psi _+$ can be related to a  
Schr\"odinger-like wave function $\Phi $ by the Darwin factor $D(r)$, i.e.
\begin{eqnarray}
\Psi _+ = \sqrt {D(r)}\ \Phi  \label{eq.scheq},
\end{eqnarray}
\begin{eqnarray}
D(r) = \frac {E+M+S(r)-V(r)}{E+M} \label{eq.Darwin}.
\end{eqnarray}

The two-component wave function $\Phi $ is solution of a Schr\"odinger equation 
containing equivalent central and spin-orbit potentials, which are functions of 
the scalar and vector potentials $S$ and $V$ and are energy dependent.

Inserting Eq.~(\ref {eq.scheq}) into Eq.~(\ref {eq.poscom}) and using the 
relativistic normalization, we obtain the wave function for the knocked out 
nucleon 
\begin{eqnarray}
\ove \Psi _f &= &\Psi ^{\dagger }_f\ \gamma ^0 = 
\sqrt {\frac {E^{\prime }+M}{2E^{\prime }}}\ \Bigglb [\pmatrix {1 \cr
\frac {\ss \cdot \p}{C}} \sqrt {D}\ \Phi _f\Biggrb ]^{\dagger }\ \gamma ^0
 \nonumber \\
 &= &\sqrt {\frac {E^{\prime }+M}{2E^{\prime }}}\ \Phi ^{\dagger }_f\ 
\bigg (\sqrt {D}\bigg )^{\dagger } 
\ \Bigglb ( 1\ ;\ \ss \cdot \p \frac {1}{C^{\dagger }}\Biggrb)\ \gamma ^0,
\label{eq.finwf} 
\end{eqnarray}
where 
\begin{eqnarray}
C(r) = E^{\prime }+M+S(r)-V(r). \label{eq.cf}
\end{eqnarray}

If we substitute Eqs.~(\ref {eq.cc2}), (\ref {eq.bouwf}), and (\ref {eq.finwf}) 
into Eq.~(\ref {eq.relcur}), we obtain the relativistic nuclear current
\begin{eqnarray}
J^0 &=& \sqrt {\frac {E^{\prime }+M}{2E^{\prime }}}\ \left\lmoustache
 \diff \r \ \Phi ^{\dagger }_f\ 
\bigg (\sqrt {D}\bigg )^{\dagger }\ \Biggl \{F_1\ \bigg [ u_i -
{\mathrm i} \ss \cdot \dd \frac {1}{C^{\dagger }}\ v_i\bigg ]\right. \nonumber \\
&& + \frac {\kappa }{2M}F_2\ \bigg [{\mathrm i}(\ss \cdot
\dd) \frac {1}{C^{\dagger }}\ (\ss \cdot \q) \ u_i + \ss \cdot \q \ v_i\bigg ]
\Biggr \}\ \exp \{{\mathrm i}\q \cdot \r \} , \nonumber \\
\J &=& \sqrt {\frac {E^{\prime }+M}{2E^{\prime }}}\ \left\lmoustache
 \diff \r \ \Phi ^{\dagger }_f\ 
\bigg (\sqrt {D}\bigg )^{\dagger }\ \Biggl \{F_1\ \bigg [ 
-{\mathrm i}(\ss \cdot \dd)
\frac {1}{C^{\dagger }}\ \ss \ u_i + \ss \ v_i\bigg ]\right. \nonumber \\
&& + {\mathrm i}\frac {\kappa }{2M}F_2\ \bigg [\ss \times \q \ u_i + \omega 
(\ss \cdot \dd) \frac {1}{C^{\dagger }}\ \ss \ u_i\nonumber \\
&& - {\mathrm i}\omega \ss \ v_i + {\mathrm i}(\ss \cdot \dd)
\frac {1}{C^{\dagger }}\ (\ss \times \q) \ v_i\bigg ]
\Biggr \}\ \exp \{{\mathrm i}\q \cdot \r \},  \label{eq.comcor}
\end{eqnarray}
where the operator $\p$ has been replaced by the gradient $-{\mathrm i}\dd$,
which operates not only on the components of the Dirac spinor but also on 
$\exp \{{\mathrm i}\q \cdot \r \}$.


\subsection{Nonrelativistic current}

In nonrelativistic DWIA the matrix elements of the nuclear current  are  
\begin{eqnarray}
J_{nr}^{\mu } = \int \diff \r \ \Phi _f^{\dagger }(\r )\ 
 \hat {\j }_{nr}^{\mu }\ \exp \{{\mathrm i}\q \cdot \r \} \ \Phi _i(\r ),
\label{eq.nonrelcur}
\end{eqnarray} 
where the bound and scattering states are eigenfunctions of a Schr\"odinger
equation. 

In standard DWIA analyses phenomenological ingredients are usually adopted for 
$\Phi _i(\r)$ and $\Phi_ f(\r)$. In this work and in order to perform a 
consistent comparison with RDWIA calculations, we employ for $\Phi _i(\r )$ the 
upper component of the Dirac wave function $u_i$ and for $\Phi _f(\r )$ the 
Schr\"odinger-like wave function that appear in the relativistic current of 
Eq.~(\ref {eq.comcor}). 

The nuclear current operator is obtained from the Foldy-Wouthuysen reduction
of the free-nucleon Dirac current through an expansion in a power series of 
$1/M$, i.e.
\begin{eqnarray}
\hat {\j }_{nr}^{\mu } = \sum _{n=0}^N\ \hat {\j }_{(n)}^{\mu }. 
\label{eq.expnrc} \end{eqnarray}
In the program {\tt DWEEPY} the expansion is truncated at second order ($N=2$).


\subsection{Cross section and response functions }

The coincidence cross section of the unpolarized ($e ,e ^{\prime }p$) reaction 
is written in terms of four nuclear structure functions
$f_{\lambda\lambda'}$~\cite{Oxford} as
\begin{eqnarray}
\sigma _0 = \sigma _{\mathrm M}\ E' |\pf |\ \bigg \{\rho _{00}
f_{00}+  \rho _{11}f_{11}+\rho _{01}f_{01}\cos {\alpha }+
  \rho _{1-1}f_{1-1}\cos {2\alpha }\bigg \},  \label{eq.fcs}
\end{eqnarray}
where $\sigma _{\mathrm M}$ is 
the Mott cross section and $\alpha $ the out of plane angle between the
electron-scattering plane and the ($\q,\p'$)-plane. The coefficients 
$\rho_{\lambda\lambda'}$ are obtained from the components of the lepton tensor
and depend only on the electron kinematics. The structure functions 
$f_{\lambda\lambda'}$  are given by suitable combinations of the components of
the nuclear current as
\begin{eqnarray}
f_{00} &=& <J^0 \big (J^0\big )^{\dagger }>, \nonumber \\ 
f_{11} &=& <J^x \big (J^x\big )^{\dagger }> + 
              <J^y \big (J^y\big )^{\dagger }>,\nonumber \\
f_{01} &=& -2\sqrt 2\ \Re \text e\ 
              \Big [<J^x \big (J^0\big )^{\dagger }>\Big ], \nonumber \\
f_{1-1} &=& <J^y \big (J^y\big )^{\dagger }> - 
               <J^x \big (J^x\big )^{\dagger }>,\label{eq.sf}
\end{eqnarray}
where the brackets mean that the matrix elements are averaged over the initial 
and summed over the final states fulfilling energy conservation.

In the following we use a different definition of the structure functions, that
is
\begin{eqnarray}
R_{\mathrm L} &=& (2\pi )^3 f_{00}, \,\,\,\,\,\,
R_{\mathrm T} = (2\pi )^3 f_{11},\nonumber \\
R_{\mathrm{LT}} &=& \frac {(2\pi )^3}{\sqrt 2} f_{01}, \,\,\,\,\,\,
R_{\mathrm {TT}} = -(2\pi )^3 f_{1-1}. \label{eq.rfun}
\end{eqnarray}

If the electron beam is longitudinally polarized with helicity $h$, the
coincidence cross section for a knocked out nucleon with spin directed along
$\hat {\s}$ can be written as~\cite{Oxford}
\begin{eqnarray}
\sigma _{\text {h},\hat {\bf {\text s}}} =  \frac {1}{2}\ \sigma _0\ 
\bigg [1 + \P \cdot \hat {\s} + h 
\Big (A + \P ^{\prime }\cdot \hat {\s}\Big )\bigg ], \label{eq.polcs}
\end{eqnarray}
where $\sigma _0$ is the unpolarized cross section of Eq.~(\ref {eq.fcs}), $\P$ 
the induced polarization, $A$ the electron analyzing power and $\P ^{\prime }$
the polarization transfer coefficient. We choose for the polarimeter the three 
perpendicular directions: $\L$ parallel to $\pf$, $\N$ along 
$\q \times \pf $, and $\T = \N \times \L$. The corresponding polarization 
observables can be written in terms of new structure functions, which contain 
explicitly the polarization direction of the emitted nucleon. In coplanar
kinematics $(\alpha =0,\pi )$, only $P^{\text N}$, $P'^{\text L}$, and 
$P'^{\text T}$ survive~\cite{Oxford}.


\section{Relativistic effects in the $^{16}$O($\e,\e'\p$) reaction}
\label{sec.3}

The reaction $^{16}$O($e,e'p$)$^{15}$N has been chosen as a well suited
process for testing the relativistic program and investigating the differences 
with respect to the nonrelativistic program {\tt DWEEPY}~\cite{DWEEPY}. 
A large experience concerning theoretical calculations is available on this
reaction and a considerable amount of experimental data at different energies
and kinematics has been published, including polarization measurements.

The relativistic bound-state wave functions used in the calculations have been 
obtained from the program {\tt ADFX} of Ref.~\cite{pvr}, where relativistic 
Hartree-Bogoliubov equations are solved.
The model is applied in the mean-field approximation to the description of 
ground-state properties of spherical nuclei~\cite{lkr}. Sigma-meson,
omega-meson, rho-meson and photon field contribute to the interaction and
their potentials are obtained by solving self-consistently Klein-Gordon 
equations. Moreover, finite range interactions are included to describe pairing
correlations and the coupling to particle continuum states.
The corresponding wave function for the nonrelativistic calculation has been
taken as the upper component of the relativistic Dirac four-component spinor 
with the proper normalization, i.e. normalized to one in the coordinate 
and spin space. 
This is not the best choice for {\tt DWEEPY}, but a consistent use of the
theoretical ingredients is necessary to allow a clear comparison between
the two approaches. 

The relativistic nuclear current was taken as in Eq.~(\ref {eq.cc2})\cite{deF}. 
This expression is not only more fundamental than the other forms recovered from 
the Gordon decomposition, but it is also consistent with the nonrelativistic 
current used in {\tt DWEEPY}, where the relativistic corrections up to order 
$1/M^2$ are obtained from a Foldy-Wouthuysen transformation applied to the 
interaction Hamiltonian where the nuclear current has the same form  as in 
Eq.~(\ref {eq.cc2}). The Dirac and Pauli form factors $F_1$ and $F_2$ are taken 
in both calculations from Ref.~\cite{mud}. 

The outgoing-proton wave function is calculated by means of the relativistic 
energy-dependent optical potential of Ref.~\cite{chc}, which fits proton 
elastic scattering data on several nuclei in an energy range up to 1040 MeV. 
The program {\tt GLOBAL} of Ref.~\cite{chc}, which generates the scalar and vector 
components of the Dirac optical potential, has been used. Different fits, 
available from the code, were explored. 
The Schr\"odinger equivalent potentials calculated in the same program
were used for the nonrelativistic calculation.

The comparison between the results of the two approaches is not restricted
to the cross section, but  involves also the structure functions, which can be 
experimentally separated and show a different sensitivity to the treatment of 
the theoretical ingredients and to relativistic effects.

A kinematics with constant ($\q$,$\omega$), or perpendicular kinematics, was 
chosen as convenient for the comparison, but some calculations were performed 
also in parallel kinematics. In both kinematics the incident electron energy 
and the outgoing proton energy are fixed. In the kinematics with constant 
($\q$,$\omega$) the electron scattering angle is calculated by imposing the 
condition $|\q|$~=~$|\p'|$. Changing the angle of the outgoing proton, different 
values of the recoil or missing momentum $p_{\mathrm m}$, with 
$\p_{\mathrm m}= \q-\p'$, can be explored. In parallel kinematics $\q$ is 
parallel or antiparallel to $\p'$, and different values of $p_{\mathrm m}$ are 
obtained changing the electron scattering angle, and thus $\q$. 

The beam energy $E_0$ was fixed in the present work at 2 GeV, in order to 
minimize the effect of the Coulomb distortion, which is included in the 
relativistic program only through the effective momentum 
approximation~\cite{Oxford,DWEEPY}. Calculations were performed for an outgoing 
proton energy up to 400 Mev. The missing momentum was explored in a range up to 
$\sim$400 MeV/$c$.

\subsection{Relativistic {\v\s} nonrelativistic results}

In this Section the results of the comparison between the DWIA calculations
performed with {\tt DWEEPY} and RDWIA calculations are discussed. 
{\tt DWEEPY} is based on a nonrelativistic treatment, but does already contain 
some relativistic corrections in the kinematics and in the nuclear current
through the expansion in $1/M$. Therefore, the results of {\tt DWEEPY} cannot be
obtained from the relativistic program simply by dropping relativistic effects,
such as the lower components of the Dirac spinor and applying the proper 
normalizations. Here the comparison is done between the results of the two
independent programs. In the first place we checked the numerical consistency of 
the two programs and verified that they give the same result in the same 
situation, i.e. when all the differences are eliminated. This numerical check 
gives us confidence that we are not interpreting the contribution of different 
ingredients as a relativistic effect.

The comparison between the results of the new RDWIA program and {\tt DWEEPY} is
shown in Figs. 1--6, for the structure functions and the cross section of 
the $^{16}$O($e,e'p$)$^{15}$N$_{\mathrm {g.s.}}$ reaction in a kinematics with 
constant ($\q$,$\omega$),  at three values of the proton energy, i.e. 
$T_{\mathrm p}$~=~100~MeV, 200~MeV and 300~MeV. 

It is clear  from these figures that the differences rapidly increase 
with the energy. Moreover, the relativistic result is smaller than the 
nonrelativistic one. This is a well known effect, which was found in all the 
relativistic calculations and which is essentially due to the Darwin 
factor~\cite{Cannata} of Eq.~(\ref {eq.Darwin}). In addition, the relativistic 
calculations include  the typical normalization factor $(E+M)/2E$ (see  
Eq.~(\ref {eq.finwf})), which has the value 0.95 at 100 MeV and decreases to 
0.87 at 300 MeV.

A significant difference is found for the transverse structure function 
$R_{\mathrm T}$ even at $T_{\mathrm p}$~=~100~MeV, where a reduction of about 
15\% is obtained with respect to the nonrelativistic calculation. The reduction
grows up to about 25\% at 200 MeV, and 40\% at 300 MeV. The 
difference is sensibly reduced, mainly at lower energies, by including in the 
nuclear current the terms to the order $1/M^3$. 

Only small differences are found for the longitudinal structure function 
$R_{\mathrm L}$ at all the considered proton energies. Its size, however, 
decreases when the energy increases and therefore its contribution to the
cross section becomes less important.

Large differences are generally found for the interference structure
function $R_{{\mathrm LT}}$. The combined relativistic effects on $R_{\mathrm T}$
and $R_{{\mathrm LT}}$ are responsible for the different asymmetry in the 
cross section at positive and negative values of $p_{\mathrm m}$, where the 
mismatch between the two structure functions adds up on the one side and 
compensates on the other side.

Large relativistic effects are also found on $R_{{\mathrm TT}}$, which is 
anyhow much smaller than the other structure functions and gives only a 
negligible contribution to the cross section.

As relativistic effects increase with the energy, the conclusion of this 
comparison is that {\tt DWEEPY} can be used with enough confidence at energies 
around 100 MeV, and, with some caution, up to about 200 MeV. Higher-order terms 
in the nuclear current can account for a part of the difference in this 
kinematical region. A fully relativistic calculation is anyhow convenient at 
200~MeV and necessary above 300~MeV. This result is consistent with the old one 
of Ref.~\cite{GP}, where the validity of the relativistic corrections to the 
nuclear current, calculated as an expansion on $1/M$, were discussed and an 
upper limit of $|\q | \sim$~600 MeV/$c$ was stated for the nonrelativistic 
calculations.

A calculation was performed also in parallel kinematics at 
$T_{\mathrm p}$~=~100 and 200~MeV. In this kinematics only the longitudinal 
$R_{\mathrm L}$ and the transverse structure function $R_{\mathrm T}$ survive. 
The results are shown in Figs.~7 and 8. Also in this case small differences are
found for $R_{\mathrm L}$, while $R_{\mathrm T}$ is significantly reduced in
RDWIA. The reduction increases with the outgoing proton energy, and, at a given 
values of $T_{\mathrm p}$, is larger at positive $p_{\mathrm m}$, when the
momentum transfer decreases. This is mainly due to the fact that the leading
term of the spin current, proportional to $\q$, is the same in the 
relativistic and nonrelativistic expressions.

\subsection{Darwin factor and spinor distortion of the scattering state}

In this Section we shall discuss the effect of the optical potential in the
Pauli reduction of the four-component Dirac spinor for the scattering state.
We do not discuss the corresponding effect on the bound state, as in our 
calculations the bound-state wave function is taken directly from the 
solution of the relativistic Dirac equation.

The effect of the optical potential on the Pauli reduction is twofold: the
Darwin factor $D(r)$ of Eq.~(\ref {eq.Darwin}), which directly multiplies the 
Schr\"odinger-equivalent eigenfunction, and the spinor distortion $C(r)$ of 
Eq.~(\ref {eq.cf}), which applies only to the lower component of the Dirac 
spinor.
The distortion of the scattering wavefunction, which is calculated through 
a partial wave
expansion, is always included in the calculation and affects in a similar way
the relativistic and nonrelativistic result.

The Darwin factor  gives a reduction of about 5-10\% at 100 MeV, 
which is consistent with the qualitative prediction of Ref.~\cite{Cannata}.
On the contrary, spinor distortion produces an enhancement of the cross
section, so that the combined effect of the two corrections is in general 
small. Moreover, it decreases with the energy. 
The result is qualitatively in agreement with the ones of 
Ref.~\cite{Ud1,Kelly3}.

In Figs. 9 and 10 the comparison between the two results is shown in the 
kinematics with constant ($\q,\omega$)  at $T_{\mathrm p}$~=~100 MeV, for the 
structure functions and the cross section, respectively. The effect is always 
small. Spinor distortion enhances the cross section at high and negative recoil 
momenta, but this effect seems absent at $p_{\mathrm m} > 0$ or it is pushed to 
very high momenta.

The effect is larger on $R_{\mathrm L}$ than on $R_{\mathrm T}$. Thus, at higher
energies, where $R_{\mathrm T}$ becomes larger than $R_{\mathrm L}$, the effect of
spinor distortion on the cross section decreases.

\subsection{Effective momentum approximation}

In this Section the validity of EMA is discussed. This prescription, which 
consists in evaluating the momentum operator in the effective nuclear current 
using the asymptotic value of the outgoing proton momentum, simplifies 
considerably the numerical calculations, avoiding the evaluation of the gradient 
in Eq.~(\ref {eq.comcor}). It is exact in plane wave impulse approximation 
(PWIA), where the scattering wave functions are eigenfunctions of the momentum, 
but in DWIA it disregards the spreading of the distorted proton wave function in 
momentum space due to FSI.

This approximation was used in some relativistic calculations, and in particular 
in the model of Refs.~\cite{Kelly2,Kelly3} for bound and scattering states. 
Since in our approach the bound-state wave function is taken directly from the 
solution of the Dirac equation, we investigate the validity of EMA only for the 
scattering state. 

We have to notice that in our calculations EMA does not change the nuclear 
current operator, which is calculated with the {\em cc2} formula and 
therefore does not depend on the momentum. 
The only dependence is contained in the Pauli reduction of the
scattering wave function.

The effect of EMA in our RDWIA approach is shown in Figs. 9 and 10, where the 
the structure functions and the cross section calculated with EMA in the 
kinematics with constant ($\q,\omega$)  at $T_{\mathrm p}$ = 100 MeV are 
displayed and compared with the exact result. At 100 MeV the difference is 
indeed sensible, but it rapidly decreases with the energy. At 200 MeV it is 
much smaller, and becomes really negligible at $\sim$ 400 MeV, and therefore at 
the energy of the TJNAF experiment. This behaviour can be understood if one 
considers that distortion effects decrease with the energy, so that at high 
energy DWIA approaches the PWIA result, where EMA is exact.


\section{Comparison with experimental data}

In this Section we shall discuss the comparison of our RWDIA results  with 
experimental data. Data are available at low energies, where {\tt DWEEPY} was 
extensively and successfully applied, and, more recently, at higher energies, 
where other RDWIA calculations have given an excellent agreement. Even though a precise
description  of experimental data is not the main aim of this work, it is 
interesting to test the predictions of our relativistic approach in comparison 
with data that have been already and successfully described by other theoretical
models. 


\subsection{ The $^{16}$O($\e,\e'\p$) reaction}

Low-energy data are presented in terms of the reduced cross
section~\cite{Oxford}, which is defined as the cross section divided by a 
kinematical factor and the elementary off-shell electron-proton scattering 
cross section, usually $\sigma_{cc1}$ of Ref.~\cite{deF}.

In Fig.\ \ref {fig.leu} the reduced cross sections measured at 
NIKHEF~\cite{Leus} for the $^{16}$O($e,e'p$) knockout reaction and for the 
transitions to the $1/2^{-}$ ground state and the $3/2^{-}$ excited state of 
$^{15}$N are displayed  and compared with the results given by our RDWIA program
and by {\tt DWEEPY}. The experiment was carried out in parallel kinematics at
$T_{\mathrm p}$~=~90 MeV. 
  
The relativistic results are lower than the nonrelativistic ones and the 
corresponding spectroscopic factors are approximately 10\% larger than those 
deduced from nonrelativistic analyses. Thus, the normalization (spectroscopic) 
factor, applied in Fig.~\ref {fig.leu} to the calculated results in order to 
reproduce the size of the experimental data, is 0.70  for RDWIA and 0.65 for 
{\tt DWEEPY}. The same value is adopted for the two final states. 

As we already stated in Sec.\ \ref {sec.3}, only small differences are found 
at this proton energy between the two models. Thus, the results of the two 
calculations are almost equivalent in comparison with data, which are reasonably
described by both calculations.  A better agreement is found for the $1/2^{-}$ 
than for the $3/2^{-}$ state. In any case, it is not as good as in the DWIA 
analysis of Ref.~\cite{Leus} performed with {\tt DWEEPY}. This result is
expected, as we already claimed that the theoretical ingredients, bound state 
and optical potential, used in the present calculation do not represent the best 
choice for {\tt DWEEPY}, but are here adopted in order to allow a consistent 
comparison between the relativistic and nonrelativtsic approaches. 
 
In Figs.\ \ref {fig.gaocs1}--\ref {fig.gaor3} the cross sections and the 
structure functions measured at TJNAF~\cite{Gao} for the $^{16}$O($e,e'p$) 
knockout reaction and for the transitions to the $1/2^{-}$ ground state and the 
$3/2^{-}$ excited state of $^{15}$N are displayed  and compared with the results 
of our RDWIA model. The experiment was carried out in a kinematics with 
constant ($\q,\omega$),  with $E_0 =$ 2.4 GeV and $\omega \sim $ 439 MeV. Only 
RDWIA calculations are shown in the figure, since we know from the 
investigation of Sec.\ \ref {sec.3} that at the proton energy of this 
experiment relativistic effects are large and  a relativistic analysis is 
necessary. In order to study the sensitivity to different optical potentials, 
we compare in the figures results obtained with the EDAD1 and EDAI-O fits of 
Ref.~\cite{chc}. Only small differences are given by the two optical 
potentials. The agreement with data is satisfactory and of about the same 
quality as in other RDWIA analyses, but for the interference structure function 
$R_{{\mathrm LT}}$ for the $1/2^{-}$ state at intermediate missing momenta. A 
better description of data might be obtained with a more careful determination 
of the theoretical ingredients.  

It is interesting to notice that the spectroscopic factor  applied to all 
the calculatations in Figs.\ \ref {fig.gaocs1}--\ref {fig.gaor3} is the same, 
i.e. 0.7, as that found in the comparison with NIKHEF data shown in 
Fig.\ \ref {fig.leu}. A spectroscopic factor of about 0.7 was also obtained in
previous RDWIA  analyses of the same TJNAF data~\cite{Gao}. 


\subsection{The $^{12}$C($\e, \e'\vec {\p}$) and 
$^{16}$O($\vec {\e}, \e'\vec {\p}$) reactions} 

The induced polarization of the outgoing proton $P^{\mathrm N}$ was measured at
 Bates for
the $^{12}$C($e,e'{\vec p}$) reaction~\cite{Woo}. Data were taken in a 
kinematics with constant ($\q,\omega$) at  $E_0 =$ 579 MeV and 
$\omega \sim $ 290 MeV. In Fig.\ \ref {fig.bates} these data are displayed and 
compared with our RDWIA calculations. Results obtained with the EDAD1 and EDAI-C 
optical potentials are compared in the figure. The EDAD1 
curve gives a better description of the experimental data at high values of the
missing momentum, but both calculations are in fair agreement with data. 
With EDAD1 potential, we also plot results after
eliminating the negative energy components in the bound state. 

As already shown in Ref.~\cite{Ud1} the polarization $P^{\mathrm N}$ is enhanced
by the presence of the negative energy components of the relativistic bound
state wave function. This result is confirmed in Fig.\ \ref {fig.bates} by the
dashed line which gives a smaller $P^{\mathrm N}$.

A slightly higher polarization is obtained in Ref.~\cite{Ud1} with the nuclear
current written in the {\em cc1} form according to Ref.~\cite{deF}.

The components of the polarization coefficient $P'^{\text L}$ and 
$P'^{\mathrm S}$ were measured 
at TJNAF~\cite{Malov} for the $^{16}$O(${\vec e},e'{\vec p}$) reaction and for 
the transitions to the $1/2^{-}$ ground state and the $3/2^{-}$ excited state 
of $^{15}$N. The experiment was performed in the same kinematics as in 
Ref.~\cite{Gao}, that is the one of Figs.\ \ref {fig.gaocs1}--\ref {fig.gaor3}. 
The experimental data are compared with our RDWIA calculations in 
Fig.\ \ref {fig.plt1} for $1/2^{-}$and in Fig.\ \ref {fig.plt3} for the 
$3/2^{-}$ state. The two curves in the figures show the results obtained 
with the EDAD1 and EDAI-O fits. Both results are in 
satisfactory agreement with data. 

\section{Summary and conclusions}

In order to clarify the differences between the usual nonrelativistic approach 
and a relativistic description of exclusive $(e, e'p)$ knockout reactions, 
we have performed a fully relativistic calculation and compared it with the 
nonrelativistic results of the
{\tt DWEEPY} code, that was successfully used to analyse a large number of
experimental data. 
The transition matrix element of the nuclear current operator is written in 
RDWIA using the relativistic bound state
wave functions obtained in the framework of the mean field theory, and
the direct Pauli reduction method with scalar and vector potentials
for the ejectile wave functions. Correspondingly, the nonrelativistic DWIA
matrix elements are computed in a consistent way to allow a direct
comparison with the relativistic results.

The main aim of this paper was not to make a precise analysis of the existing
experimental data, but to discuss the use of a nonrelativistic approach at
energies higher than those generally considered up to few years ago, and to
clarify the possible relativistic effects arising also at lower energies.

We have used the new RDWIA and {\tt DWEEPY} codes to perform calculations for
several kinematical conditions. The relativistic results are always smaller than
the nonrelativistic ones and the difference increases with energy. 
The transverse and interference structure functions are particularly sensitive
to relativistic effects, much more than the longitudinal structure
function. 
$R_{\mathrm T}$ is sensibly reduced even at low energy:
inclusion of higher order terms in the nonrelativistic nuclear current
can reduce the difference, but a fully relativistic calculation is
necessary above $T_{\text p}\sim $ 200 MeV.

The effect of the scalar and vector potentials in the Pauli reduction
for the scattering state has been discussed. These potentials appear in the 
relativistic treatment and are absent in the nonrelativistic one.
The combined contribution of the reduction due to 
the Darwin factor and of spinor distortion, which enhances the
effects of the lower components of the Dirac spinor, is always small.

The validity of EMA in the scattering state of relativistic calculations 
has been studied.
The differences with respect to the exact results are sensible at 
$T_{\mathrm p}$ = 100 Mev, but rapidly decrease with 
the energy and become negligible at $T_{\text p}\sim $ 400 MeV.

We have tested our new RDWIA calculations in comparison with experimental data 
that have
already been described by other models. The agreement is satisfactory
and comparable with other relativistic analyses.


\newpage 
\begin{figure}
\begin{center}
\epsfig{file=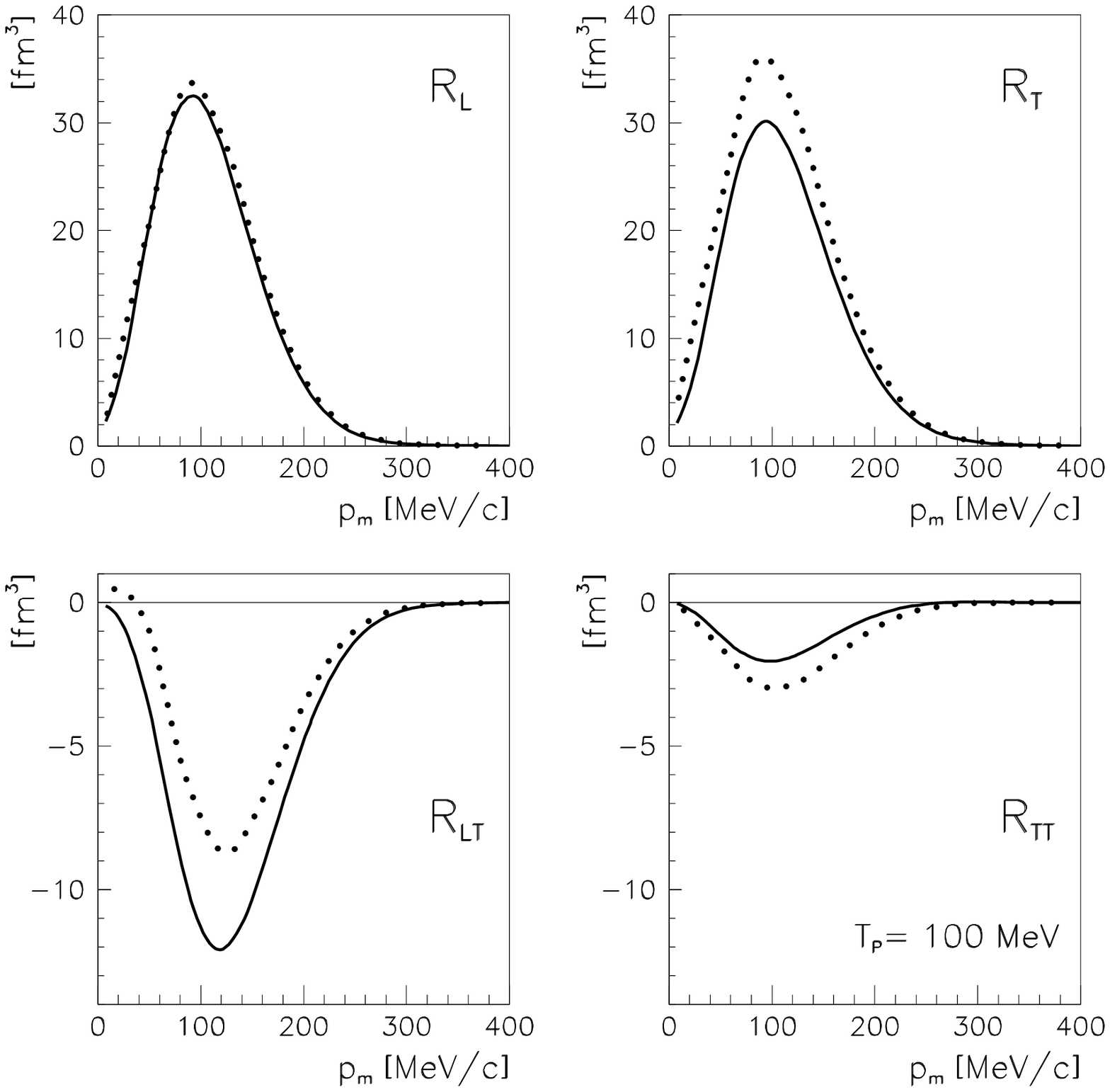,width=10cm}
\end{center}
\caption[]{The structure functions of the $^{16}$O($e,e'p$)
reaction as a function of the recoil momentum $p_{\mathrm{m}}$ for the 
transition to the $1/2^{-}$ ground state of $^{15}$N, in a kinematics with 
constant ($\q,\omega$), with $E_{0} = 2000$ MeV and $T_{\mathrm p} = 100$ MeV. 
The solid lines give the RDWIA result, the dotted lines the nonrelativistic 
result of {\tt DWEEPY}. Positive (negative) values of $p_{\mathrm m}$ refer to
situations where the angle between $\p'$ and the incident electron $\k$ is 
larger (smaller) than the angle between $\q$ and $\k$.   
\label{fig:fig1}
}
\end{figure}

\begin{figure}
\begin{center}
\epsfig{file=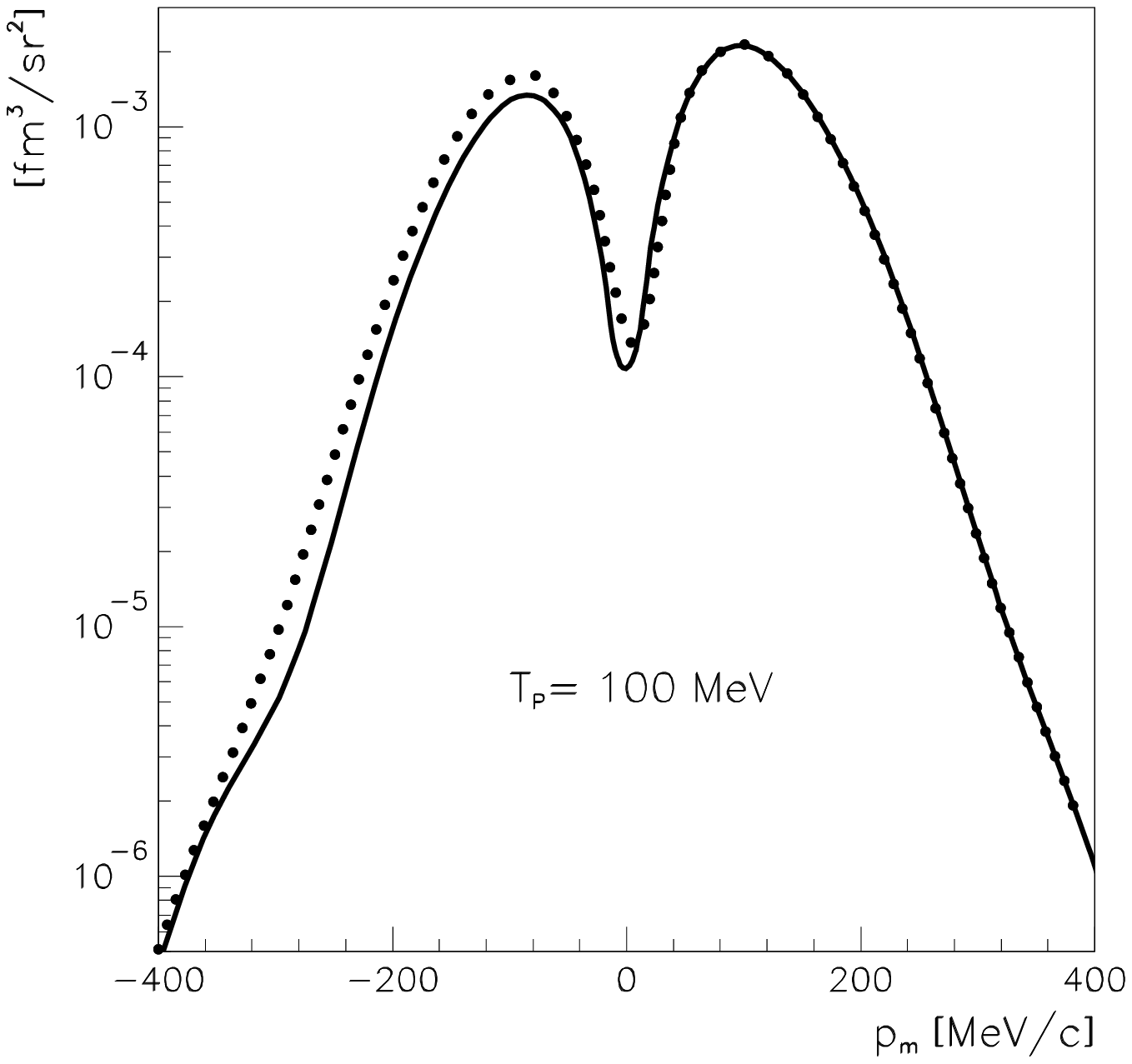,width=10cm}
\end{center}
\caption[]{The same as in Fig. 1, but for the cross section of the 
$^{16}$O($e,e'p$) reaction.
\label{fig:fig2}
}
\end{figure}

\begin{figure}
\begin{center}
\epsfig{file=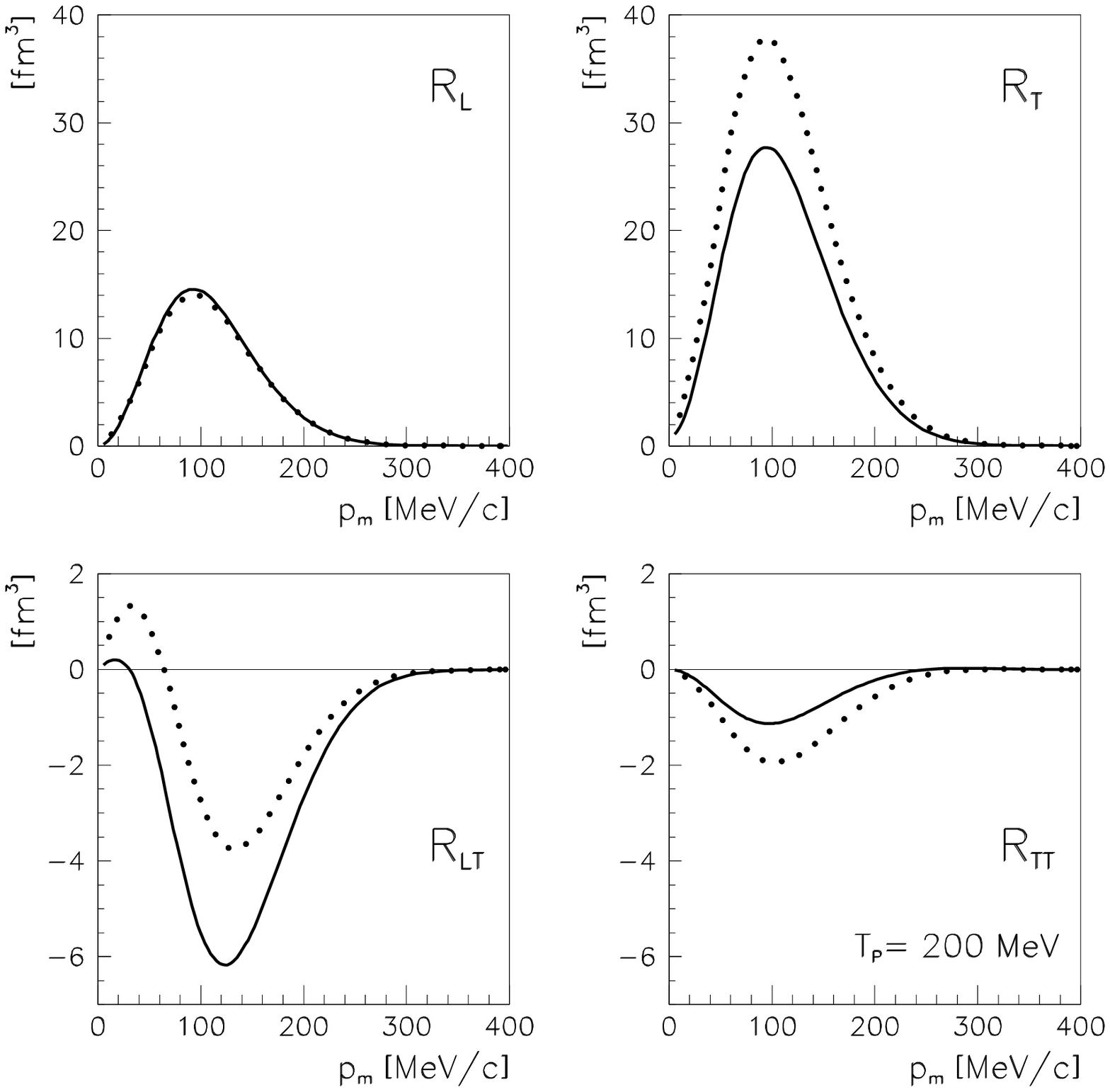,width=10cm}
\end{center}
\caption[]{The same as in Fig. 1 at $T_{\mathrm p} = 200$ MeV.
\label{fig:fig3}
}
\end{figure}

\begin{figure}
\begin{center}
\epsfig{file=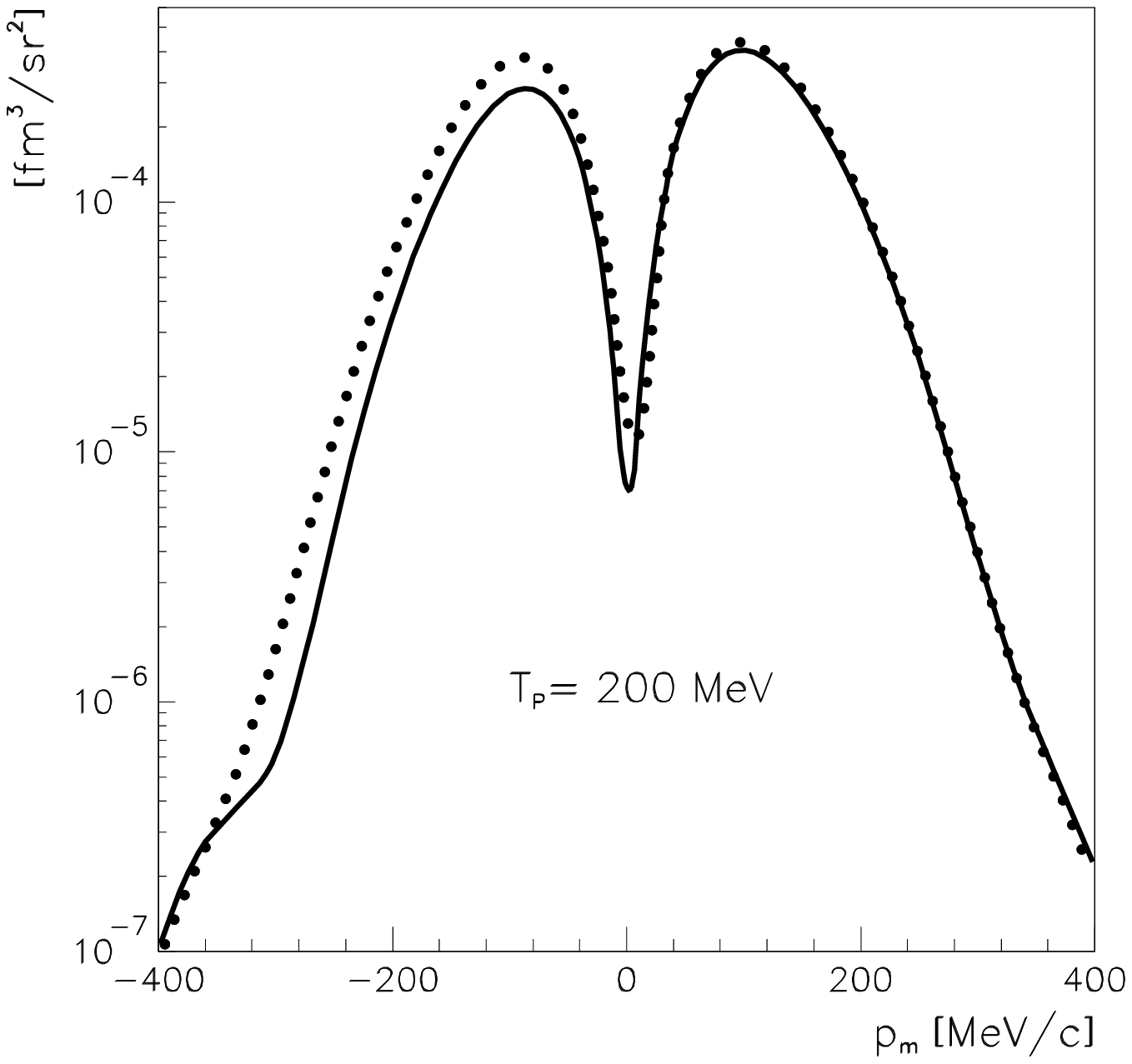,width=10cm}
\end{center}
\caption[]{The same as in Fig. 2 at $T_{\mathrm p} = 200$ MeV.
\label{fig:fig4}
}
\end{figure}

\begin{figure}
\begin{center}
\epsfig{file=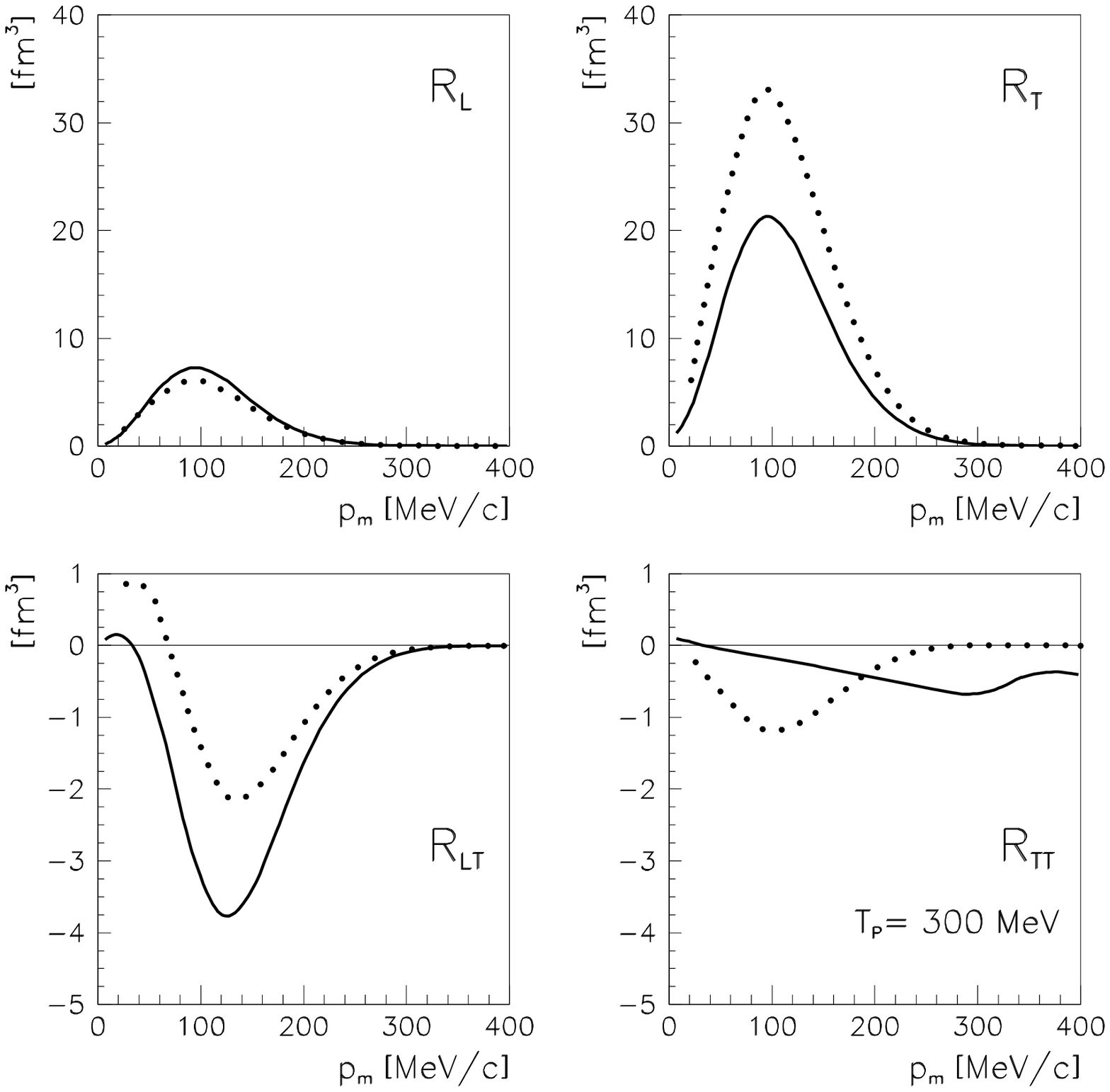,width=10cm}
\end{center}
\caption[]{The same as in Fig. 1 at $T_{\mathrm p} = 300$ MeV.
\label{fig:fig5}
}
\end{figure}

\begin{figure}
\begin{center}
\epsfig{file=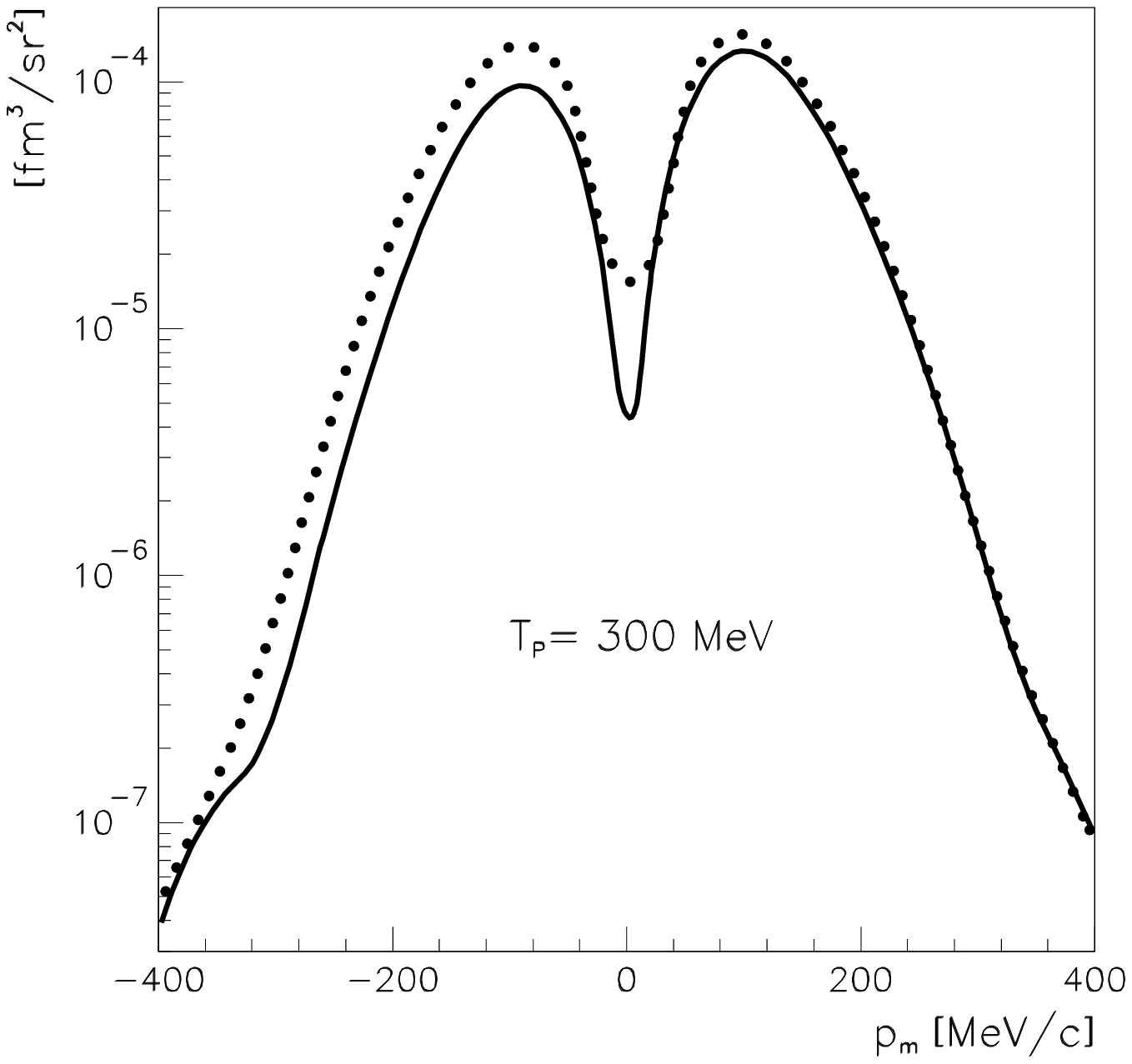,width=10cm}
\end{center}
\caption[]{The same as in Fig. 2 at $T_{\mathrm p} = 300$ MeV.
\label{fig:fig6}
}
\end{figure}

\begin{figure}
\begin{center}
\epsfig{file=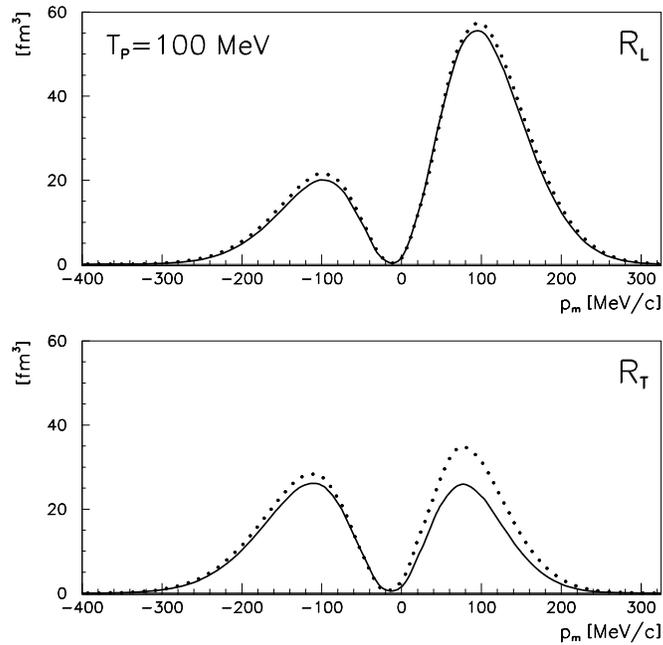,width=10cm}
\end{center}
\caption[]{The longitudinal and transverse structure functions of the 
$^{16}$O($e,e'p$) reaction as a function of the recoil momentum 
$p_{\mathrm{m}}$ for the transition to the $1/2^{-}$ ground state of $^{15}$N,
in parallel kinematics with $E_0 =$ 2000 MeV and $T_{\mathrm p} = 100$ MeV. 
Positive (negative) values of $p_{\mathrm{m}}$ refer to situations where 
$|\p'| > |\q|$ ($|\p'| < |\q|)$.
Line convention as in Fig. 1.
\label{fig:fig7}
}
\end{figure}

\begin{figure}
\begin{center}
\epsfig{file=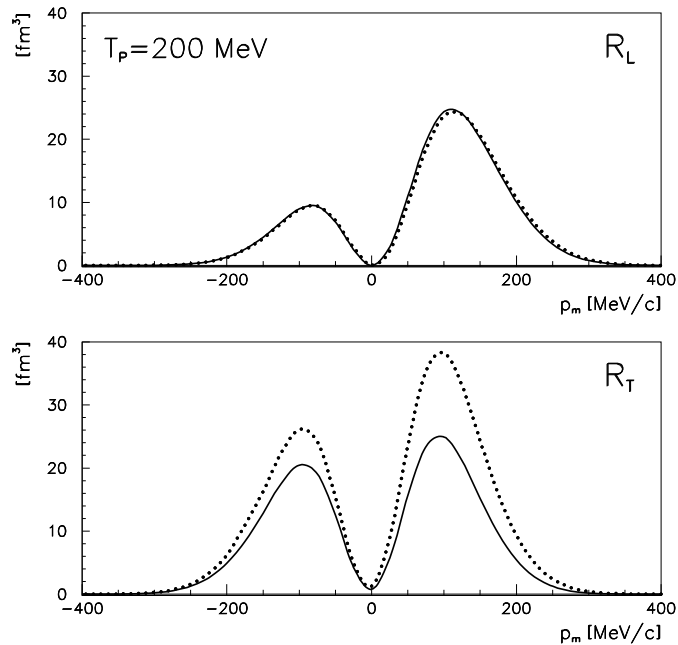,width=10cm}
\end{center}
\caption[]{The same as in Fig. 7 for $T_{\mathrm p} = 200$ MeV.
\label{fig:fig8}
}
\end{figure}
 
\begin{figure}
\begin{center}
\epsfig{file=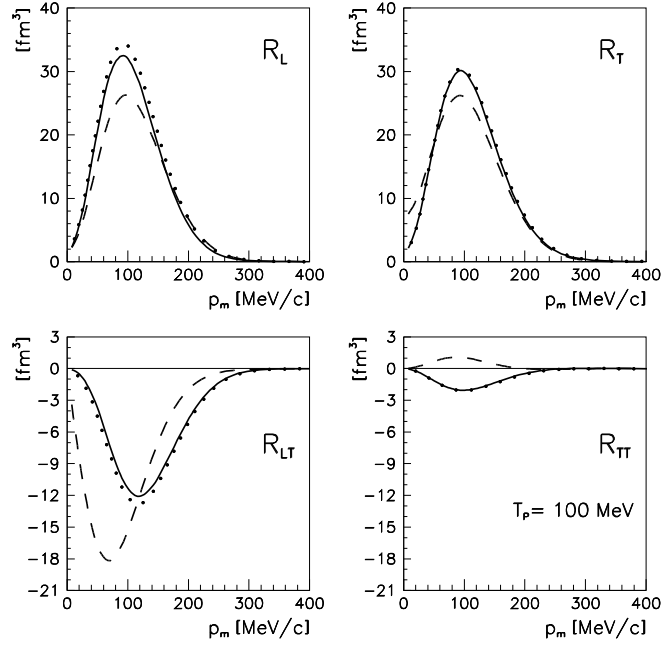,width=10cm}
\end{center}
\caption[]{The structure functions of the $^{16}$O($e,e'p$)
reaction as a function of the recoil momentum $p_{\mathrm{m}}$ for the 
transition to the $1/2^{-}$ ground state of $^{15}$N, in the same situation as 
in Fig. 1. The solid lines give the RDWIA result, the dotted lines the
calculation without the Darwin factor and spinor distortion and the dashed 
line the EMA.  
\label{fig:fig9}
}
\end{figure}

\begin{figure}
\begin{center}
\epsfig{file=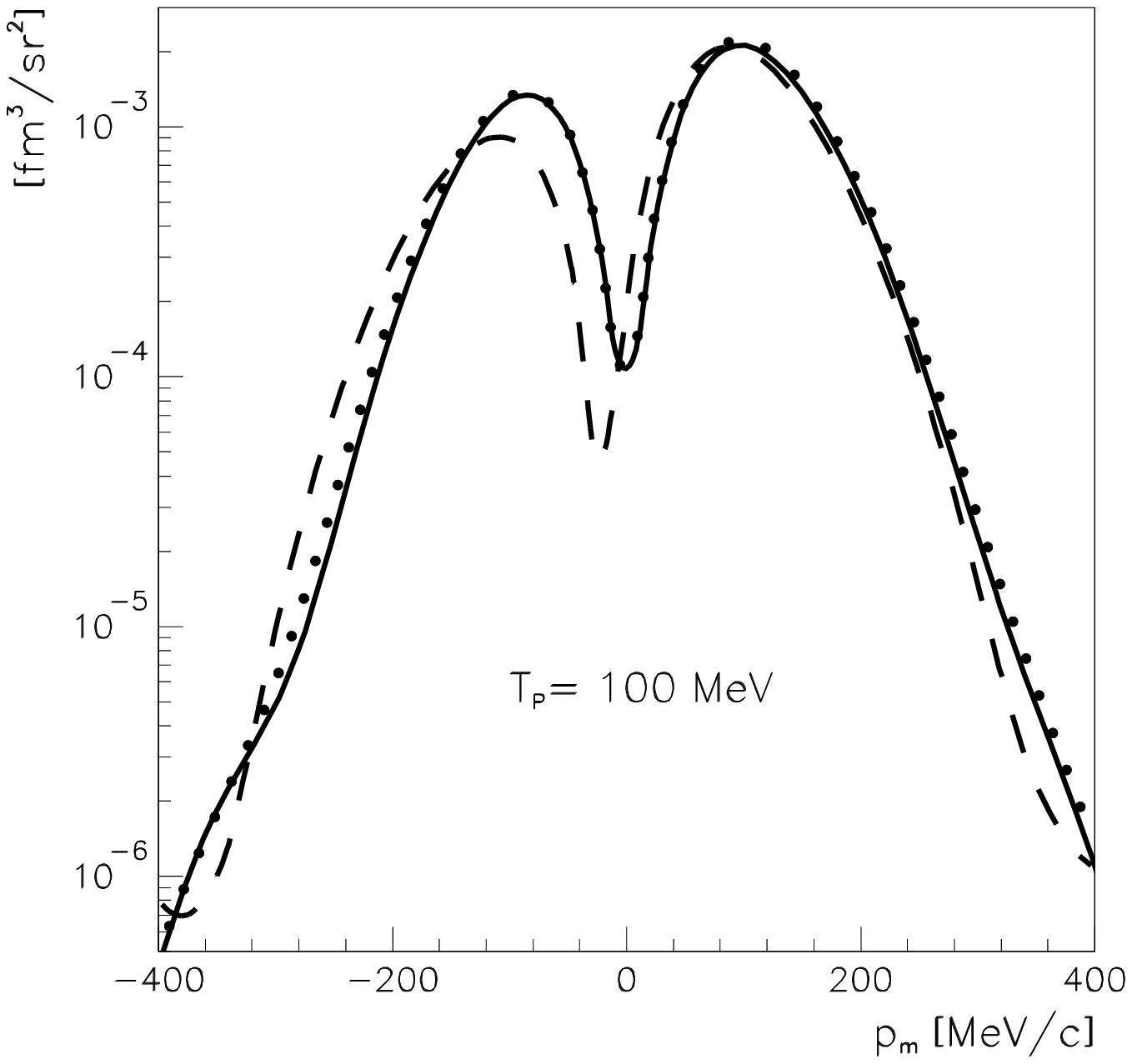,width=10cm}
\end{center}
\caption[]{The same as in Fig. 9, but for the cross section of the 
$^{16}$O($e,e'p$) reaction.
\label{fig:fig10}
}
\end{figure}

\begin{figure}
\begin{center}
\epsfig{file=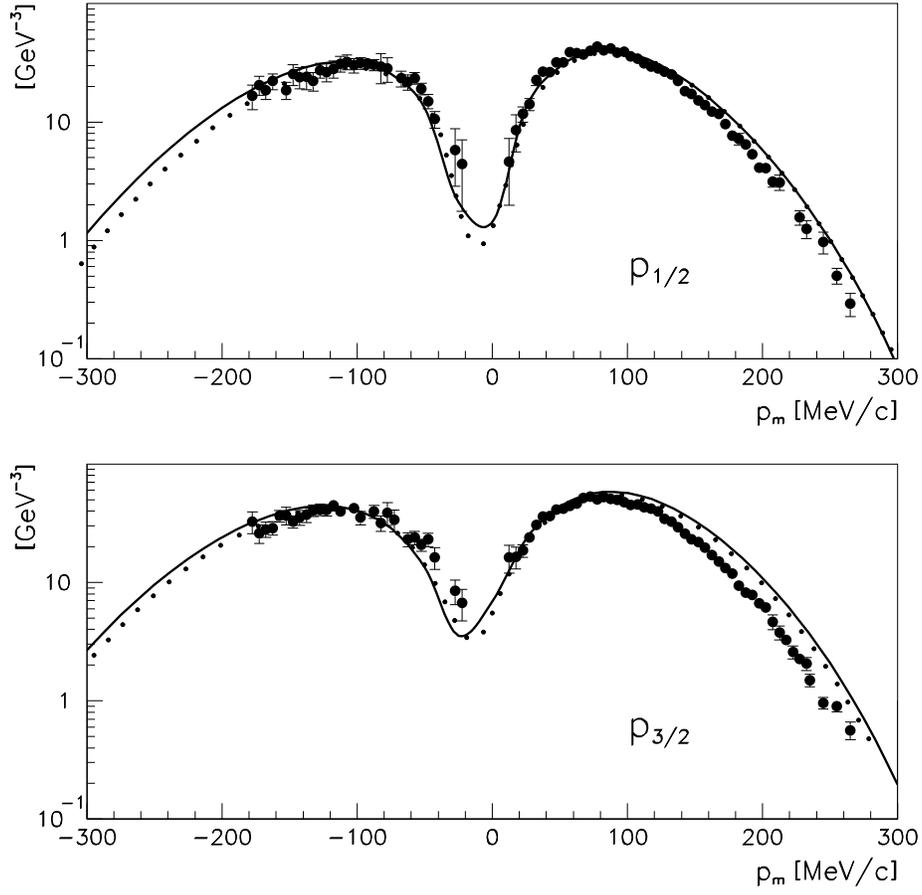,width=14cm}
\end{center}
\caption[]{The reduced cross section of the $^{16}$O($e,e'p$) reaction as a 
function of the recoil momentum $p_{\mathrm{m}}$ for the transitions to the 
$1/2^{-}$ ground state and to the $3/2^{-}$ excited state of $^{15}$N, in 
parallel kinematics with $E_0 =$ 520 MeV and $T_{\mathrm p} =$ 90 MeV. The data 
are from Ref. \cite {Leus}. The solid lines give the RDWIA result, the dotted 
lines the nonrelativistic result of {\tt DWEEPY}.}                
\label{fig.leu}
\end{figure}

\begin{figure}
\begin{center}
\epsfig{file=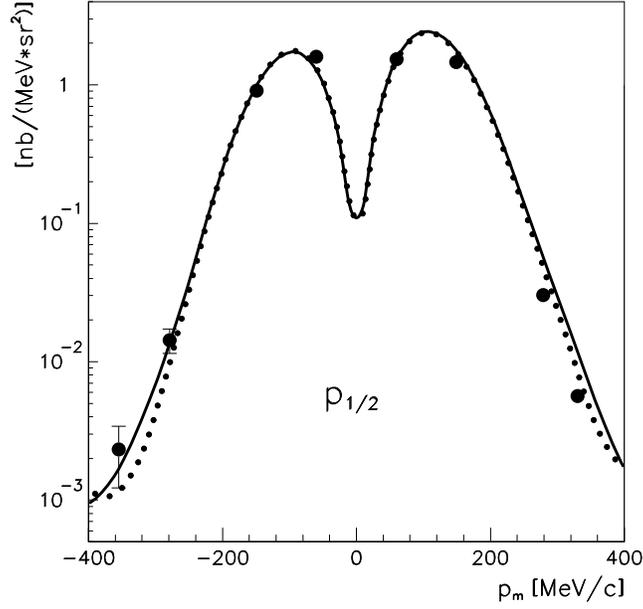,width=10cm}
\end{center}
\caption[]{The cross section of the $^{16}$O($e,e'p$) reaction as a 
function of the recoil momentum $p_{\mathrm{m}}$ for the transition to the 
$1/2^{-}$ ground state of $^{15}$N in a kinematics with constant ($\q,\omega$), 
with $E_{0} = 2445$ MeV and $T_{\mathrm p} = 433$ MeV. The data are from 
Ref. \cite {Gao}. The solid line gives the RDWIA result with the EDAD1 
optical potential, the dotted line the RDWIA result with the EDAI-O optical 
potential. }                
\label{fig.gaocs1}
\end{figure}
\begin{figure}
\begin{center}
\epsfig{file=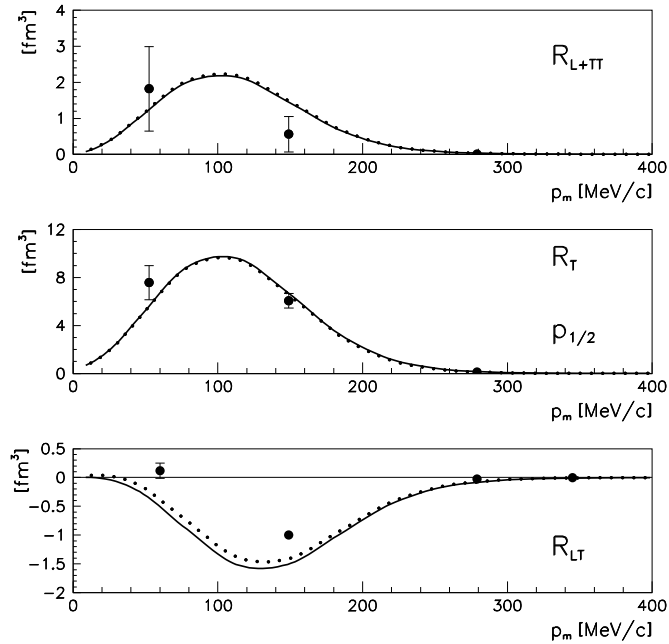,width=10cm}
\end{center}
\caption[]{The same as in Fig.\ \ref {fig.gaocs1} but for the response 
functions of the 
$^{16}$O($e,e'p$) reaction. }                
\label{fig.gaor1}
\end{figure}
\begin{figure}
\begin{center}
\epsfig{file=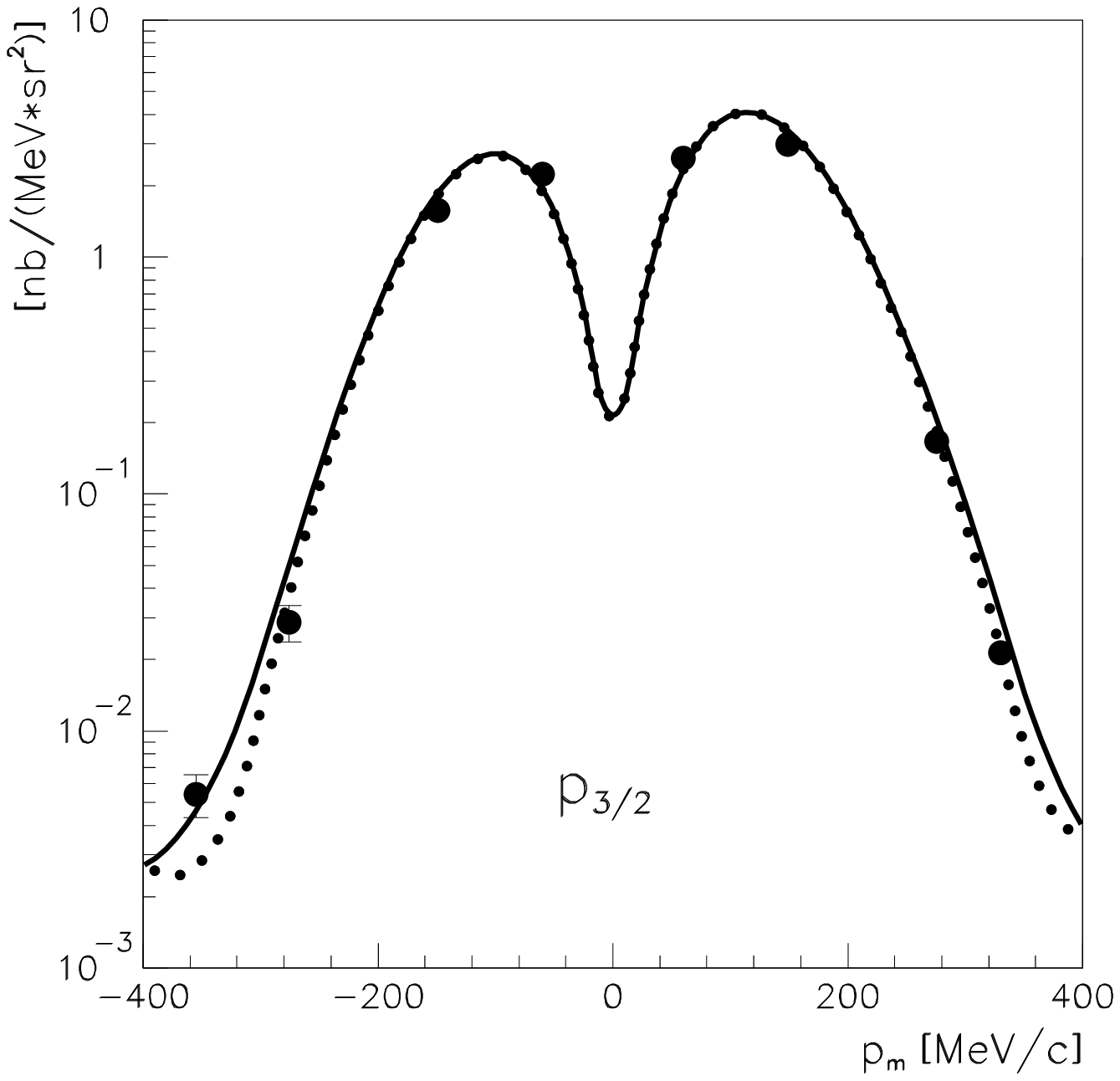,width=10cm}
\end{center}
\caption[]{The same as in  Fig.\ \ref{fig.gaocs1} but for the transition to the 
$3/2^{-}$ excited state of $^{15}$N. }                
\label{fig.gaocs3}
\end{figure}
\begin{figure}
\begin{center}
\epsfig{file=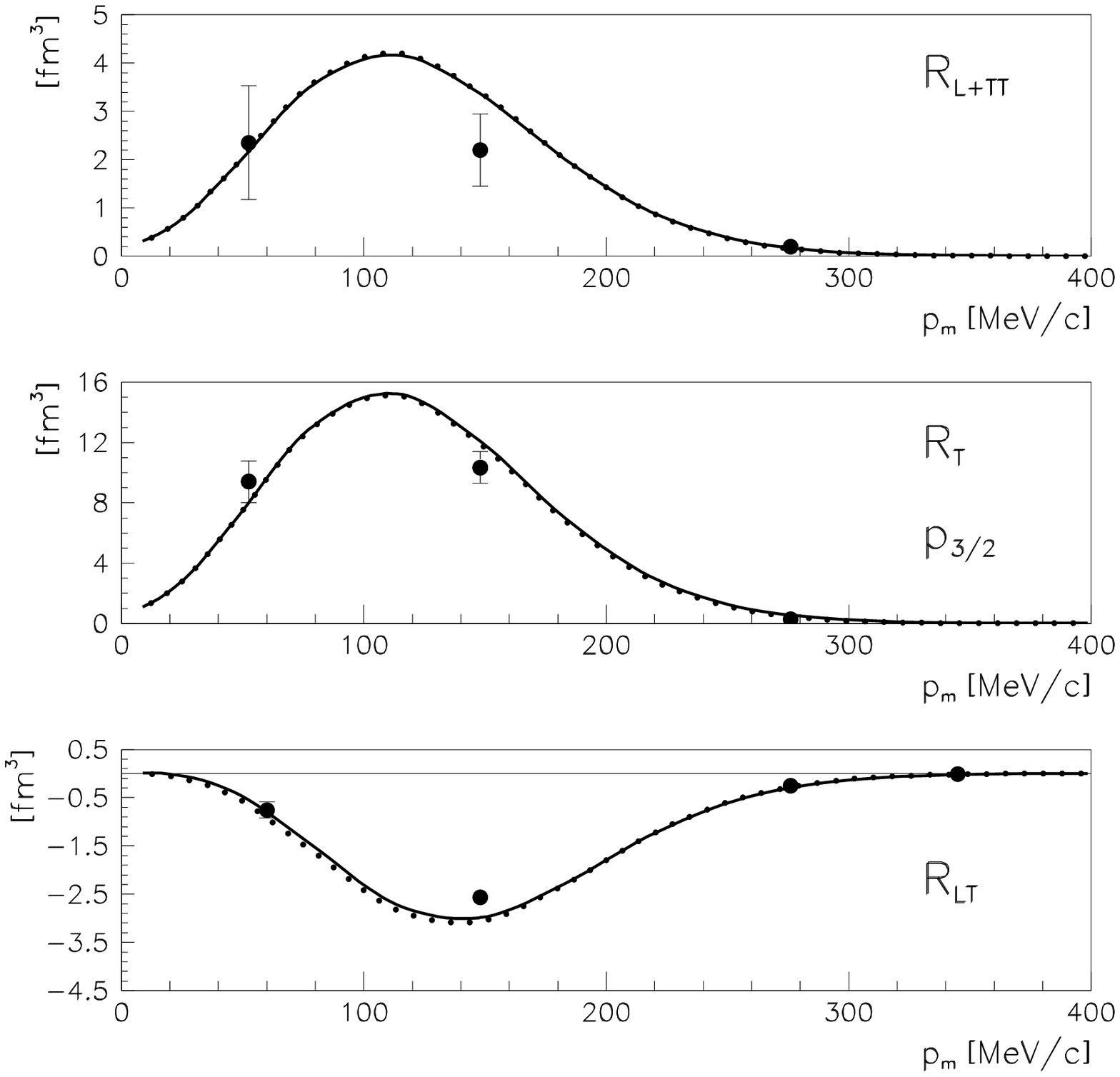,width=10cm}
\end{center}
\caption[]{The same as in  Fig.\ \ref{fig.gaor1} but for the transition to the 
$3/2^{-}$ excited state of $^{15}$N. }                
\label{fig.gaor3}
\end{figure}
\begin{figure}
\begin{center}
\epsfig{file=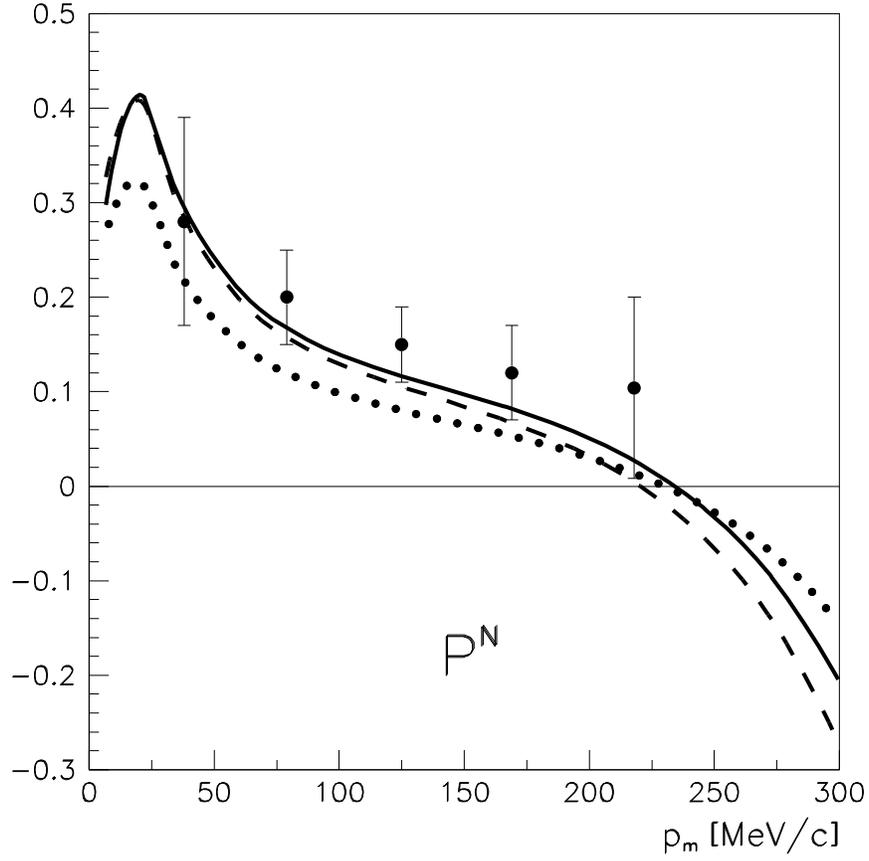,width=14cm}
\end{center}
\caption[]{The induced polarization of the emitted proton for the 
$^{12}$C($e, e'\vec p)$ reaction as a function of the recoil momentum 
$p_{\mathrm{m}}$ for the transition to the $3/2^{-}$ ground state of $^{11}$B 
in a kinematics with constant ($\q,\omega$), with $E_{0} = 579$ MeV and 
$T_{\mathrm p} = 274$ MeV. The data are from Ref. \cite {Woo}. The solid line
gives the RDWIA result with the EDAD1 optical potential, the dotted line the 
RDWIA result with the EDAI-C  optical potential, and the dashed line the result
with the EDAD1 optical potential and after removing the negative energy 
components in the bound state.}                
\label{fig.bates}
\end{figure}
\begin{figure}
\begin{center}
\epsfig{file=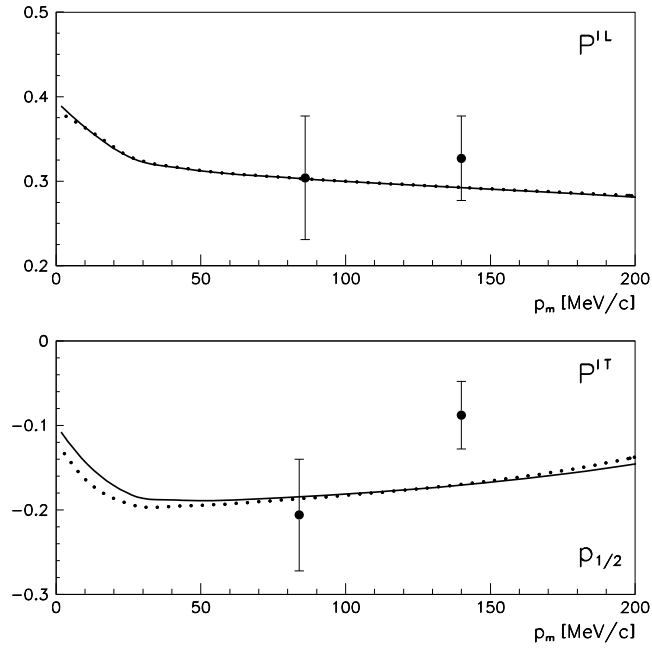,width=10cm}
\end{center}
\caption[]{The components of the polarization transfer coeffcient $P'^{\text L}$
and $P'^{\text T}$ for the $^{16}$O($\vec e, e'\vec p$) reaction as a function 
of the recoil momentum $p_{\mathrm{m}}$ for the transition to the $1/2^{-}$ 
ground 
state of $^{15}$N in the same kinematics as in Fig.\ \ref {fig.gaocs1}. 
The data are from Ref. \cite {Malov}. The solid lines give the RDWIA result 
with the EDAD1 optical potential, the  dotted lines the RDWIA result with 
the EDAI-O optical potential.}                
\label{fig.plt1}
\end{figure}
\begin{figure}
\begin{center}
\epsfig{file=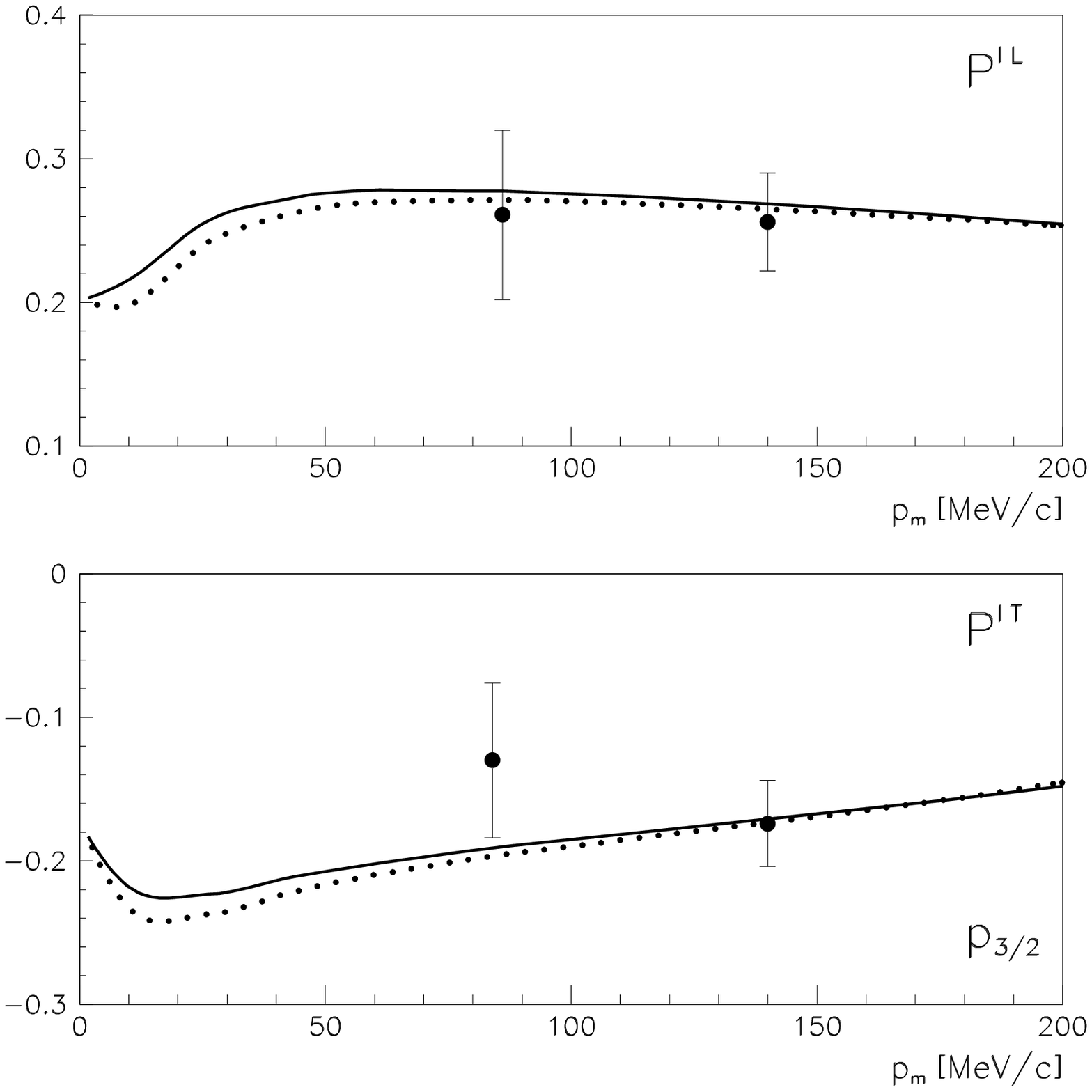,width=10cm}
\end{center}
\caption[]{The same as in  Fig.\ \ref{fig.plt1} but for the 
transition to the $3/2^{-}$ excited state of $^{15}$N.}                
\label{fig.plt3}
\end{figure}

\end{document}